\documentclass[aps,prl,twocolumn,groupedaddress,superscriptaddress,showpacs]{revtex4-1}

\usepackage{graphicx}
\usepackage{color}
\usepackage{amsmath}
\usepackage{amsfonts}
\usepackage{amssymb}
\usepackage{dcolumn}
\usepackage{booktabs}

\usepackage{hyperref}
\hypersetup{colorlinks=true,urlcolor=blue,linkcolor=blue,citecolor=blue}
\usepackage[final]{pdfpages}

\begin{document}

\title{Breathing synchronization in interconnected networks}

  \author{V. H. P. Louzada}
    \affiliation{Computational Physics, IfB, ETH-Honggerberg, Wolfgang-Pauli-Strasse 27, 8093 Zurich, Switzerland}

  \author{N. A. M. Ara\'ujo \footnote{Correspondence and requests for materials should be addressed to N. A. M. A. (nuno@ethz.ch)}}
    \affiliation{Computational Physics, IfB, ETH-Honggerberg, Wolfgang-Pauli-Strasse 27, 8093 Zurich, Switzerland}

  \author{J. S. Andrade, Jr.}
    \affiliation{Computational Physics, IfB, ETH-Honggerberg, Wolfgang-Pauli-Strasse 27, 8093 Zurich, Switzerland}
    \affiliation{Departamento de F\'isica, Universidade Federal do Cear\'a, 60451-970 Fortaleza, Cear\'a, Brazil}

  \author{H. J. Herrmann}
    \affiliation{Computational Physics, IfB, ETH-Honggerberg, Wolfgang-Pauli-Strasse 27, 8093 Zurich, Switzerland}
    \affiliation{Departamento de F\'isica, Universidade Federal do Cear\'a, 60451-970 Fortaleza, Cear\'a, Brazil}
 
\date{\today}

\begin{abstract}
Global synchronization in a complex network of oscillators emerges from
the interplay between its topology and the dynamics of the pairwise
interactions among its numerous components. When oscillators are
spatially separated, however, a time delay appears in the interaction
which might obstruct synchronization.  Here we study the synchronization
properties of interconnected networks of oscillators with a time delay
between networks and analyze the dynamics as a function of the couplings
and communication lag.  We discover a new breathing synchronization
regime, where two groups appear in each network synchronized at
different frequencies. Each group has a counterpart in the opposite
network, one group is in phase and the other in anti-phase with their
counterpart. For strong couplings, instead, networks are internally
synchronized but a phase shift between them might occur. The
implications of our findings on several socio-technical and biological
systems are discussed.
\end{abstract}

\maketitle

Technology has furnished us with global connectivity changing the
functioning of cooperative work, international business, and
interpersonal relationships. For example, despite the ever faster
Internet connections, there will always be a physical limit speed to
information transport, thereby imposing a time delay in communication.
As we discuss here, this time delay might pose a real challenge to the
synchronizability of oscillators. Therefore, understanding the
consequences of a communication lag is of major concern in different
fields~\cite{Duke2006,Boccaletti2006,Helbing2009}.  For example, the
plasmodium \emph{Physarum polycephalum}, an amoeba-like organism
consisting of a network of tubular structures for protoplasm flow,
naturally shows periodic variations in its thickness, a necessary
feature for its survival. A controlled setup has been prepared by
Takamatsu {\it et al.} where two regions of the same organism have been
physically separated by a certain distance with the possibility of fine
tuning the communication between them
\cite{Takamatsu2000,Takamatsu2009}. Depending on the coupling strength
and time delay, the two regions have been shown to present phase and
anti-phase synchronization of the oscillatory thickness. This is
precisely what we find in the regime of strong intra-network coupling.
As discussed in the final section, this biological system might be a
prototype to experimentally evaluate the different regimes reported
here. In what follows, we discuss the general case of two interconnected
networks but our study might have impact on several biological and
techno-social systems as, for example, functional brain networks, living
oscillators, or coupled power grids, as discussed at the end of this
paper. 

Recent geometrical studies of coupled networks with intra- and
inter-network links have revealed novel features never observed for
isolated networks~\cite{Gao2011}. In particular, it has been shown that
the overall robustness is
reduced~\cite{Schneider2011,Schneider2013,Herrmann2011,Schneider2013b,Louzada2013a}
and the collapse of the system occurs through large cascades of
failures~\cite{Buldyrev2010,Brummitt2011}. Dynamic properties of coupled
networks have also been
studied~\cite{Li2007,Sorrentino2007,Wu2009,Shang2009,Mao2012,Araujo2013,Cardillo2013,Gomez2013},
but the impact of a time delay on their synchronization is still an open
issue, which we will address here. Typically, the intra- and
inter-network couplings have different time scales. For simplicity, we
consider the case where intra-network interactions can be considered
instantaneous and the inter-network ones have a communication lag that
depends on the distance between networks. In particular, we show that,
when isolated, the two networks would naturally move in unison. However,
when interacting the oscillators in the same network split into two
groups, synchronized with different frequencies, leading to breathing
synchronization.

The Kuramoto model is the standard theoretical framework for studying
synchronizability of
networks~\cite{Kuramoto1987,Neda2000,Boccaletti2002,Wang2002a,Pikovsky2003,Strogatz2003,Li2004,Lu2005,Motter2005,Acebron2005,Osipov2007,Arenas2008,Boccaletti2008,Barthelemy2008,Louzada2012,Nicosia2013}.
A population $\Theta$ of $n$ Kuramoto oscillators is considered to be
mutually interacting. We consider a random graph of average degree four.
Each oscillator $i\in\Theta$ is described by a phase $\theta_i(t)$,
representing its current position, and a natural frequency $\omega_i$.
For simplicity, we assume the same frequency $\omega_i\equiv\omega_0$
for all oscillators. The actual frequency of an oscillator is defined as
the time derivative of the phase, $\dot{\theta_i}(t)$. To move
harmoniously, oscillators try to synchronize their frequencies and
phases. This interaction can be modeled in terms of the Kuramoto model
as
$\dot{\theta}_i=\omega_0+\sigma\sum_{j=1}^{n}A_{ij}^{\Theta}\sin\left(\theta_j-\theta_i\right)$,
where the sum goes over all other oscillators ($i\neq j$), $\sigma$ is
the coupling strength between them,  and $\mathbf{A^{\Theta}}$ is the
connectivity matrix such that $A_{ij}^{\Theta}=1$ if oscillator $i$ is
influenced by $j$ and zero otherwise. We also assume that all
oscillators have the same unitary amplitude, so that the state of each
one can be described by a phasor $e^{i\theta_i(t)}$. 

The collective motion, namely, the synchronization of the network, is
characterized here by the complex order parameter ${r_\Theta(t)
e^{i\Psi(t)}= \frac{1}{n}\sum_{j=1}^n e^{i\theta_j(t)}}$, where the sum goes
over all oscillators, $\Psi(t)$ is the average phase, and the amplitude
${0\leq|r_\Theta(t)|\leq1}$ measures the global coherence, i.e., how
synchronized the oscillators are. If $r_\Theta(t)=1$ all oscillators are
synchronized, while very low values of $r_\Theta$ imply that a
significant fraction of oscillators are out of phase.

We introduce now a second population $\Gamma \neq \Theta$, also of $n$ oscillators interacting in a random graph of average degree four,
representing the second network. We couple each $j \in \Gamma$ with one, and only one,
corresponding partner $i \in \Theta$, forming the inter-network
couplings. In analogy to oscillators in $\Theta$, the motion of each
oscillator is described by a phasor $e^{i\gamma_j(t)}$, of phase
$\gamma_j(t)$. The inter-network coupling is subjected to a time delay
$\tau$, corresponding to the time required for information to travel
between networks~\cite{Schuster1989}. Previous studies introduced time
delay among oscillators of the same
population~\cite{Yeung1999,Choi2000}. Here we consider the competition
between an \emph{instantaneous} intra-network and a \emph{delayed}
inter-network coupling. In a nutshell, the dynamics of oscillators is described by,
\begin{equation}
\scalebox{0.87}{ \begin{minipage}{\columnwidth}
$\left\{
\begin{array}{l}
\dot{\theta}_i=\omega_0+ {\sigma_{\text{EX}}} \sin\left(\gamma^{t-\tau}_{j(i)} - \theta_i \right)+ \sigma_{\text{IN}}\displaystyle\sum_{k=1}^{N}A_{ik}^{\Theta}\sin\left(\theta_k-\theta_i\right) \\
\dot{\gamma}_j=\omega_0+ {\sigma_{\text{EX}}} \sin\left(\theta^{t-\tau}_{i(j)} - \gamma_j \right)+ \sigma_{\text{IN}}\displaystyle\sum_{k=1}^{N}A_{ik}^{\Gamma}\sin\left(\gamma_k-\gamma_j\right)
\end{array} \right.$
\end{minipage} },
\label{eq::time_delay}
\end{equation} 
where the superscript $t-\tau$ indicates the instant when the phases are
calculated, and $\sigma_{\text{EX}}$ and $\sigma_{\text{IN}}$ are the
inter and intra-network couplings, respectively. 

\begin{figure}[h]
\includegraphics[width=\columnwidth]{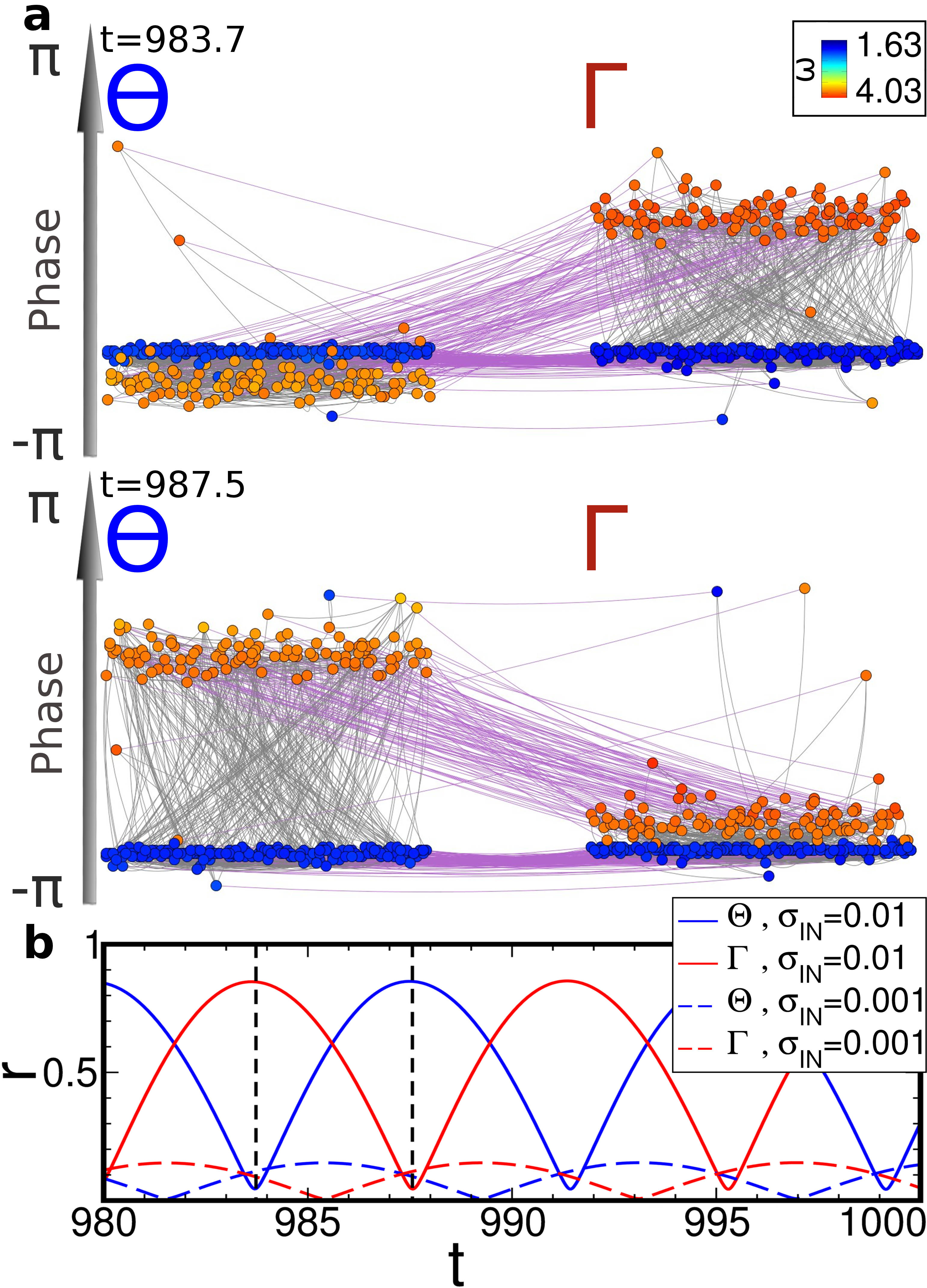}
\caption{\textbf{The interactions between a strongly delayed inter-network coupling and a weak intra-network coupling create two communities of different frequencies in steady state}. \textbf{a}, Snapshot of populations at two different time steps (black dashed vertical lines in b) near the steady state, for $\omega_0=1.0$, $\tau=1.53$, $\sigma_{\text{IN}}=0.01$, and $\sigma_{\text{EX}}=0.5$. The vertical position of each oscillator represents its phase, from $-\pi$ to $\pi$, and the color represents the frequencies achieved with oscillators mostly presenting values near the theoretical frequencies (1.63 and 4.63) of the steady state. Superposition of these two communities leads to breathing synchronization. \textbf{b}, Time evolution of the order-parameter of populations $\Theta$ (blue) and $\Gamma$ (red) composed of $n=305$ oscillators each with $\omega_0=1.0$, $\tau=1.53$, and $\sigma_{\text{EX}}=0.5$. Two scenarios of weak intra-network coupling are represented: $\sigma_{\text{IN}}=0.
01$ (continuous lines) and $\sigma_{\text{IN}}=0.001$ (dashed lines).}
\label{fig::breathing_r}
\end{figure}

\begin{figure*}[!ht]
\includegraphics[width=\textwidth]{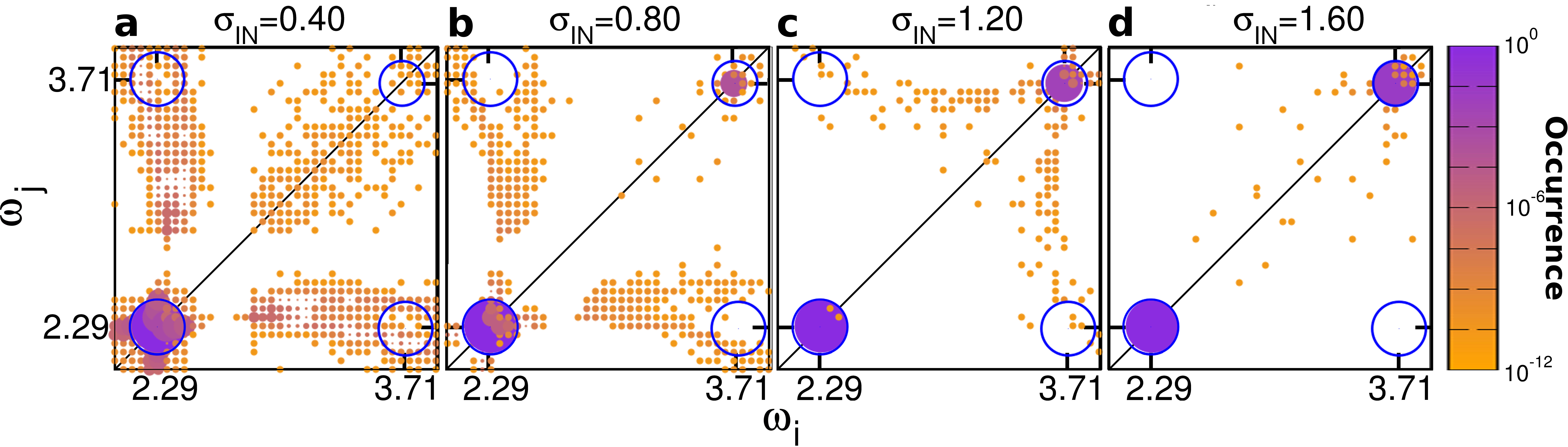}
\caption{\textbf{Scatter plot for the matrix of frequency pairs of intra-network neighboring oscillators}, for $500$ different realizations of random coupled networks of $n=750$, with $\omega_0=2.75$, $\tau=1.53$, $\sigma_{\text{EX}}=1.5$, and various $\sigma_{\text{IN}}$, namely, $0.4$ (\textbf{a}), $0.8$ (\textbf{b}), $1.2$ (\textbf{c}), and $1.6$ (\textbf{d}). Empty and filled circles are centered on the frequency pairs $(\omega_i,\omega_j)$ for each $(i,j)$ neighboring nodes within a network, calculated with a 2D binning of size $0.05$. The filled circles color, according to a purple-yellow scale, corresponds to their relative occurrence in the dataset: purple circles are the predominant frequencies registered, while yellow circles are less common. For comparison, blue empty circles correspond to results for $\sigma_{\text{IN}}=0$. The size of the symbols is also used for the relative occurrence of each pair.}
\label{fig::wi_x_wj}
\end{figure*}

\section*{Results}

We observe that for two interconnected networks of oscillators with time
delay, a weak intra-network coupling, and random initial distribution of
phases, two frequency communities emerge within the same network, each
synchronized with its mirror in a breathing mode, as shown in
Fig.~\ref{fig::breathing_r}(a). In the figure, the color describes the
frequency and the vertical position the phase. The frequency
synchronization within groups occurs with phase locking. Interestingly,
inter-network coupled pairs of nodes oscillate with the same frequency
(same color) but might be either in phase or anti-phase (phase shift of
$\pi$). Consequently, the presence of these two frequency groups affects
the perception of the new global oscillatory state, which we call
breathing synchronization. Figure~\ref{fig::breathing_r}(b) shows the
time evolution of the order parameters $r_\Theta$ and $r_\Gamma$ for
each population, quantifying this breathing behavior. For each curve,
the maximum corresponds to the instant at which both groups of
frequencies are in phase, while the minimum to an anti-phase between
groups in the same network. Additionally, since for one frequency there
is a phase shift of $\pi$ between inter-network pairs of nodes, the
minimum in one network corresponds, necessarily, to the maximum in the
other.  Cohesion within each community affects the amplitude of the
breathing, as indicated by the order parameters for different values of
$\sigma_{\text{IN}}$ in Fig.~\ref{fig::breathing_r}(b). The weaker the
intra-network coupling, the smaller is this amplitude. 

The observed breathing behavior is in deep contrast with what is
expected for an isolated network ($\sigma_{\text{EX}}=0$). For isolated
networks, the classical Kuramoto model is recovered, with frequency and
phase synchronization emerging at a critical coupling
$\sigma_{\text{IN}}=\sigma_{\text{IN}}^*$. Above this threshold, a
macroscopic fraction of oscillators is synchronized, all with the same
frequency and phase. The value of $\sigma_{\text{IN}}^*$ increases with
the variance of the natural frequency distribution. Since here we
consider the same natural frequency for all oscillators
($\omega_i\equiv\omega_0$), $\sigma_{\text{IN}}^*\rightarrow 0$. The
group of synchronized oscillators has frequency
$\omega=\omega_0$ and the order parameter
$r_\Theta(t)$ (or $r_\Gamma(t)$) saturates in time at a non-zero
steady-state value~\cite{Kuramoto1987}, which is a monotonically increasing function of
($\sigma_{\text{IN}}-\sigma_{\text{IN}}^*$). Interestingly, in the case
of coupled networks, and for sufficient inter-network couplings, none of
the two frequencies is $\omega_0$.

To better understand the breathing synchronization, and in particular
the emergence of frequency groups, let us consider the case of two
coupled oscillators with time delay. The analytic solution obtained by
Schuster and Wagner~\cite{Schuster1989} for this problem indicates that,
depending on the initial phase difference between oscillators (see
Fig.~S1 in the Supplemental Material), the pair can
synchronize with different frequencies $\omega$, which are solutions of,
\begin{equation}
 \omega = \omega_0 - \sigma_{\text{EX}} \sin\left(\omega\tau\right).
\label{eq:schuster}
\end{equation}
In spite of oscillating with the same frequency in the stationary state,
the two oscillators might either be in phase, if $\cos(\omega\tau)>0$,
or anti-phase, otherwise. In the case of inter-connected networks, in
the limit $\sigma_{\text{IN}}=0$, the stationary state is expected to
include all possible solutions of Eq.~\ref{eq:schuster}. Surprisingly,
our results with a weak coupling reveal instead two frequency groups
with phase locking. Nevertheless, the observed frequencies are
consistent with the solution of Eq.~\ref{eq:schuster} and are unique with respect to $\omega_0$ and $\tau$. The final frequency of a pair of oscillators only depends of their relative initial displacement.

As we show next, when the internal coupling ($\sigma_{\text{IN}}$) is
further increased, breathing synchronization is no longer stable and
each network is synchronized, in one of two other synchronization
regimes. 
In simulations with $\sigma_{\text{EX}}=1.5$, $\omega_0=2.75$ and $\tau=1.53$, when $\sigma_{\text{IN}}=0.4$ interactions among oscillators in
the same network become more relevant than the inter-network delayed coupling, and the larger frequency group, in terms of size, dominates over the smaller one. This competition results in all oscillators synchronizing at the same frequency and the order parameter of each network saturates in time. 
To systematically study the dependence on $\sigma_{\text{IN}}$,
we analyze the frequency correlation among intra-network neighbors $i$
and $j$. Figure~\ref{fig::wi_x_wj} shows the scatter plots of the pair
($\omega_i,\omega_j$) for different values of intra-network coupling
strengths. The limit $\sigma_{\text{IN}}=0$ is represented by the blue
empty circles in all panels and the radius corresponds to the relative
population of pairs when considering several samples. In this limit, the
oscillators have all one of two possible frequencies, with four possible
combinations of frequency pairs. From the relative size of the circles,
we observe that the lowest frequency ($\omega\approx2.3$ for
$\omega_0=2.75$ and $\tau=1.53$) is the most populated one.
As shown in Fig.~\ref{fig::wi_x_wj}(a), for
$\sigma_{\text{IN}}=0.4$ most nodes are synchronized with the lowest
frequency and therefore a large percentage of the pairs are in the
left-bottom corner. Similarly to the Kuramoto model, in this
\emph{competing} state, oscillators synchronize at a unique stable
frequency ($\omega\approx2.3$), which is a solution of
Eq.~\ref{eq:schuster}. As $\sigma_{\text{IN}}$ is further increased
(Fig.~\ref{fig::wi_x_wj}(b)-(d)), due to the strength of the
intra-network coupling, each network tends to behave like a
\emph{supernode} and, depending on the initial conditions, one of two
frequencies is obtained, which is again a solution of
Eq.~\ref{eq:schuster}. Further analysis across samples (See Fig.~S2 in
Supplemental Material) also shows that the average phase
displacement between pairs of oscillators in different networks reaches
$\Delta=\pi$ for intermediary values of $\sigma_{\text{IN}}$, and
decreases again once the supernodes are formed (See Fig.~S2(a) in
Supplemental Material). For large $\sigma_{\text{IN}}$, the
supernodes can be either in phase or anti-phase and, therefore, the
average variance within a network has a value between zero and $\pi$
(See Fig.~S2(b) in Supplemental Material). Results are qualitatively similar for networks with fixed node degree (See Fig.~S2(c-d) in Supplemental Material) or with different average degree (See Fig.~S2(e-f) in Supplemental Material).

\begin{figure}[!h]
\includegraphics[width=\columnwidth]{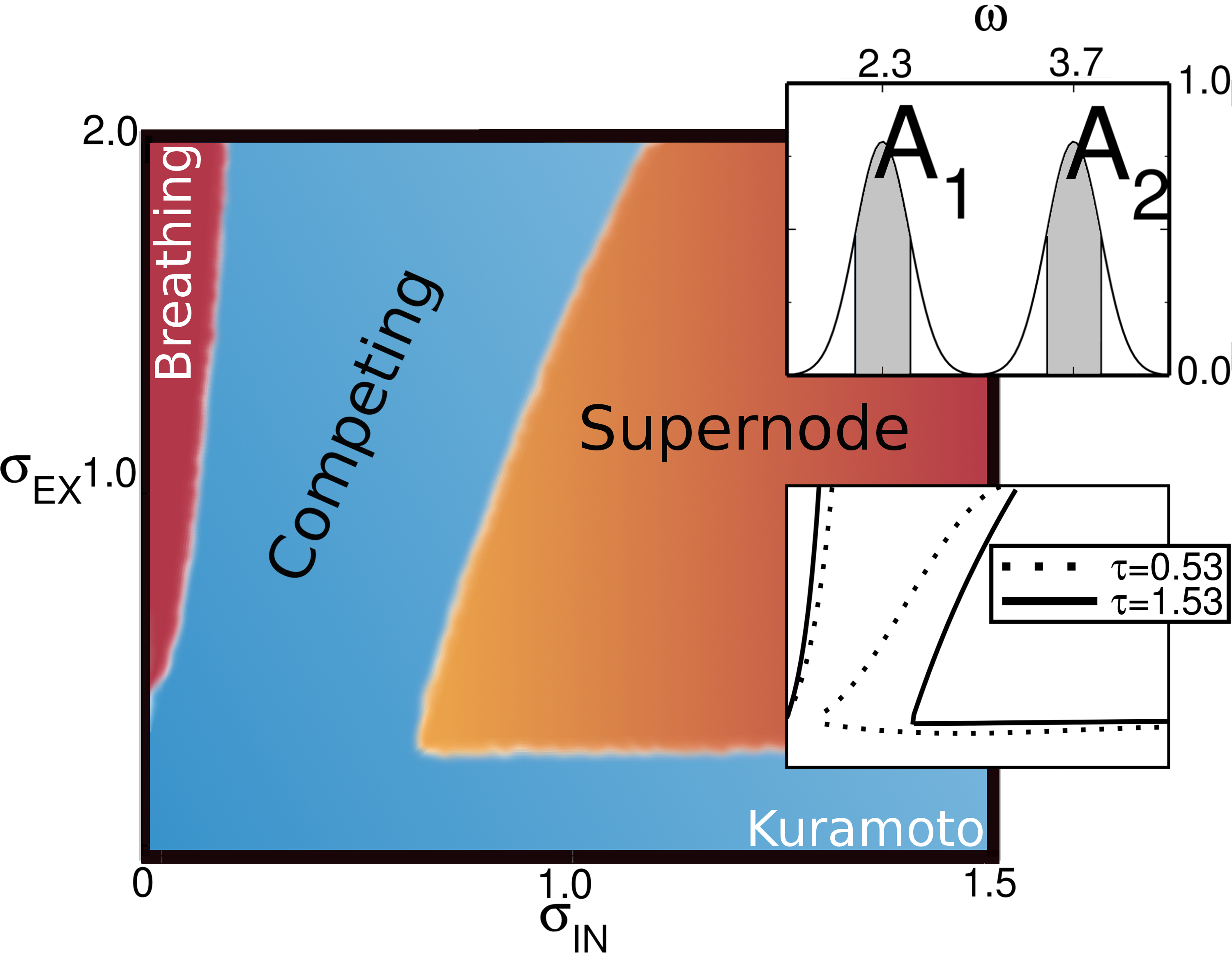}
\caption{\textbf{Phase diagram for delayed coupled networks}. Parameter space of two coupling strengths $\sigma_{\text{EX}}$ and $\sigma_{\text{IN}}$ showing that the prevalence of one frequency over the other changes according to the coupling strengths. The color of each region represents the occurrence of the two theoretical frequencies: red if two frequencies ($\omega=2.3$ and $\omega=3.7$) and blue if only one ($\omega=2.3$) is observed. Shaded regions mark the boundaries between states. Top inset is an example of the histogram used to calculate the main panel: areas around the theoretical frequencies are defined ($A_1$ and $A_2$) and their ratio used to define the prevalence of only one or two of them. The lower inset exhibits the state boundaries for different time delays. The dominant mechanisms of each region are labeled accordingly: Breathing, Kuramoto~\cite{Kuramoto1987}, Competing, and Supernode states. Regions are defined based on simulations over 300 different realizations of 
random coupled networks of $n=500$, with $\omega_0=2.75$ and $\tau=1.53$.}.
\label{fig::colormap}  
\end{figure}

To summarize the effect of several combinations of parameters, we plot
in Fig.~\ref{fig::colormap} the phase diagram in the space of the two
coupling strengths ($\sigma_{\text{IN}}$ and $\sigma_{\text{EX}}$). To
identify each regime, we compute the amount of oscillators with steady
frequency below and above the mean value of possible frequencies (see
Section Methods), $A_1$ and
$A_2$, respectively, over different samples (see top inset of
Fig.~\ref{fig::colormap}). The color map of the main plot of
Fig.~\ref{fig::colormap} shows the ratio of these quantities. While the
blue area represents the domain of $\sigma_{\text{IN}}$ and
$\sigma_{\text{EX}}$ combinations that leads to the smaller frequency,
the shades in red represent the two regions where two frequencies can be
achieved. Note that, the nature of the two synchronization regimes in
red is different. The one in the left (lower $\sigma_{\text{IN}}$) is
characterized by the breathing behavior due to the presence of two
frequency groups within each network. By contrast, in the supernode
regime all nodes within a network are in phase locking, with the same
frequency and, therefore, the order parameter is constant in time in the
steady state. In the bottom inset, we show the phase boundaries for
different time delays. From this, one can also see that the transition
between regimes changes substantially for different time delays. Since delay and natural frequency are not multiples, harmonic interactions are considered negligible. Table~\ref{tab::summary} contains a brief summary of all states reported in Fig.~\ref{fig::colormap}.
In Fig. S3 of the Supplemental Material we show that the transitions between
regimes with one and two stable frequencies are abrupt. 

\begin{center}
\begin{table*}[!ht]
\caption{\textbf{Summary of different synchronization regimes.} A brief description of the properties of all states encountered for $\omega_0=2.75$ and $\tau=1.53$.}
\begin{tabular}[c]{|c|>{\centering}m{0.5\textwidth}|c|}
\hline
\textbf{State} & \textbf{Frequency groups} & \textbf{Phase shift} \\ \hline
Breathing & Two groups: $\omega=2.29$ (low freq.) and $\omega=3.71$ (high freq.) & $\pi$ for low freq. and $0$ for high freq.\\ \hline
Competing & One group: $\omega=2.29$& $\pi$ \\ \hline
Supernode & One group: $\omega=2.29$ or $\omega=3.71$& $\pi$ if low freq. or $0$ if high freq. \\  
\hline
\end{tabular}
\label{tab::summary}
\end{table*}
\end{center}

\section*{Discussion}

The presence of a time delay between two coupled networks of oscillators
poses a new challenge to the global control of the system. We have shown
that the interplay between coupling and delay leads to states of either
a unique or two possible synchronized frequencies. We have found that,
even with a weak intra-network coupling, oscillators within the same
network split into two frequency groups. Each group has a mirror one in
the other network oscillating at the same frequency. However, depending
on their frequency, a group can be either in phase or anti-phase with
its mirror in the other network, resulting in breathing synchronization.
Also, we show that an arbitrary increase of the intra-network coupling
is not an option to achieve phase and frequency synchronization
regardless of its initial conditions. In a certain region of the
parameter space, the intra-network coupling promotes the formation of
two supernodes (one per network), and two frequencies become stable. We have numerically identified the transition regions between regimes. Future works should consider recent advances on group synchronization to analytically study these transitions through linear stability analysis using the master stability function~\cite{Dahms2012,Williams2013}.

As previously mentioned, it is possible to prepare controlled
experiments to evaluate the existence of these different regimes in
biological systems. Takamastu {\it et al.} \cite{Takamatsu2000} have
shown that the distance and interaction strength between regions of a
plasmodial slime mold can be fine tuned. This organism is a network of
tubular structures with periodic variations in the thickness. In the
experimental study, the focus was only on the regime where the
intra-region interaction is much stronger than the inter-region one.
Using the same methodology, it is possible to control the intra-region
interaction and study the different regimes described here. In
particular, it would be of interest to observe oscillations with two
different frequencies within the same region due to the communication
lag with the other region, resulting in breathing synchronization.

Another example where synchronization in interdependent networks
certainly plays a relevant role is the human brain. Being a highly
modular structure, its coherent operation must rely on the independence
of different brain modules, which are functionally specialized, as well
as on their efficient connection to ensure proper information
transmission and processing. In a recent study~\cite{Gallos2012}, it was
shown that the optimal integration of these modules, which can be
interpreted as complex networks made of intra-network couplings, is
achieved through the addition of long-range inter-network ties,
therefore behaving globally as a small-world system. Moreover, their
experimental observations are also consistent with the fact that these
inter-network couplings should be spatially organized in such a way as
to maximize information transfer under wiring cost
constraints~\cite{Li2010,Li2013}. To accomplish multisensory integration in
this intricate architecture of neuronal
firing-oscillators~\cite{Hodgkin1952}, however, information originating
from distinct sensory modalities (vision, audition, taction, etc.) must
ultimately be processed in a synchronized way. This is typically the
case when the processing of a visual signal influences the perception of
an auditory stimulus and vice-versa~\cite{Shams2000,Hairston2003}.

\section*{Methods}

Equation~\ref{eq::time_delay} has been numerically solved using a fourth order Runge-Kutta method with discrete time steps $\delta t=0.003$. The stable frequencies were computed at $t_{max}=100$, using the difference between phases after one $\delta t$ step. The natural frequency has been chosen as $\omega_0=1.00$ in Fig.~\ref{fig::breathing_r} and $\omega_0=2.75$ for Figs.~\ref{fig::wi_x_wj}-\ref{fig::colormap}. Initial phases of oscillators in all simulations have been sampled from a random uniform distribution between $-\pi$ and $\pi$. Different values of $\omega_0$ do not affect qualitatively the results. The same values of $\delta t$ and $t_{max}$ were adopted for all simulations in this study. 

In Fig.~\ref{fig::breathing_r}, Panels a) and b) are based on one pair of undirected random networks of average degree four and $305$ nodes in each. Oscillators in this figure have been simulated for $\tau=1.53$, $\sigma_{\text{EX}}=1.5$. Panel a) is based on $\sigma_{\text{IN}}=0.01$.

In Fig.~\ref{fig::wi_x_wj}, Panels a)-d) contain the simultaneous representation of 500 pairs of random networks of $750$ nodes. Color and size of each point represents the relative occurrence in all data. Oscillators in this figure have been simulated for $\tau=1.53$ and $\sigma_{\text{EX}}=1.5$. 

Fig.~\ref{fig::colormap} is a schematic representation based on the average over 300 pairs of undirected random networks of average degree four and 500 nodes in each. The upper inset is a graphical representation of the histogram of all stable frequencies. The lower inset contains the same study for different delays, also averaged over 300 pairs of undirected random networks of average degree 4 and 500 nodes in each. A cutoff of $\omega=3.00$, the midpoint of the stable frequencies for $\sigma_{\text{IN}}=0$, was used to determine the areas $A_1$ and $A_2$. Colors in the main panel are defined according to the ratio of A1 and A2: blue if $log(A_1/A_2)<4$ and red if $log(A_1/A_2)>4$, with shades of these colors used to represent the transition regions. To avoid the effect of oscillators that did not reach a stable state by the end of the simulation, we consider only frequencies with a relative occurrence of more than 10\%. Oscillators in this figure have been simulated with $\tau=1.53$ in the 
main panel and $\tau=0.53$ in the lower panel.

\clearpage
\begin{acknowledgments}
\textbf{Acknowledgments.} Authors would like to thank the Swiss National Science Foundation under contract 200021 126853, the CNPq, Conselho Nacional de Desenvolvimento Cient\'i­fico e Tecnol\'ogico - Brasil, the CNPq/FUNCAP Pronex grant, the ETH Zurich Risk Center, and the INCT-SC-Brasil for financial support. This work was also supported by grant number FP7-319968 of the European Research Council. We would like also to thank L. de Arcangelis, A. Gower, F. Mohseni, and K. J. Schrenk for the valuable discussions.
\end{acknowledgments}

\section*{Authors Contributions}
V.H.P.L., N.A.M.A., J.S.A., and H.J.H. wrote the main text, prepared the simulations, and discussed the results.

\section*{Additional Information}
\textbf{Competing financial interests:} The authors declare no competing financial interests.

\clearpage


\section*{Supplementary material}

# Supplemental Material
### Breathing synchronization in interconnected networks

V. H. P. Louzada,[1] N. A. M. Araújo,[1] J. S. Andrade, Jr.,[1,2] and H. J. Herrmann[1,2]

[1]*Computational Physics, IfB, ETH-Honggerberg, Wolfgang-Pauli-Strasse 27, 8093 Zurich, Switzerland*
[2]*Departamento de Física, Universidade Federal do Ceará, 60451-970 Fortaleza, Ceará, Brazil*

## STEADY STATE FREQUENCY ACHIEVED FOR VARYING INITIAL CONDITIONS

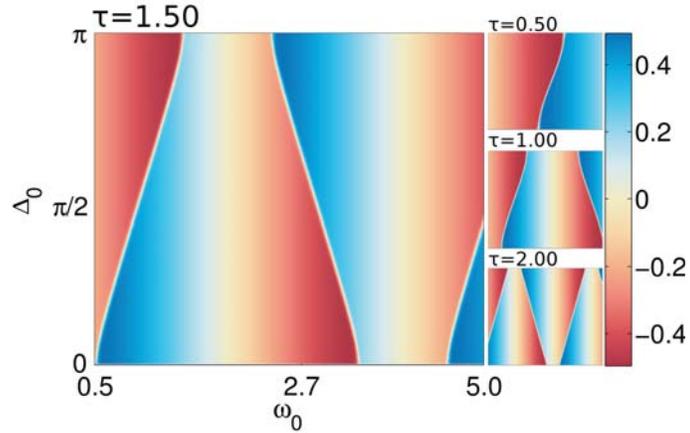

FIG. S1: For a combination of $\Delta_0$, the initial phase displacement between inter-network oscillators, and $\omega_0$, we map the final frequency $\omega$ achieved in the case of a weak intra-network coupling. The color scale represents the variable $\omega - \omega_0$. As for our simulations phases are initially randomly distributed, blue and red areas can be seen as the size of the basin of attraction of different frequency solutions. Data is an average of 500 sampels of $n = 576$ oscillators simulated with $\sigma_{\text{IN}} = 0.01$ and $\sigma_{\text{EX}} = 0.50$. Main and lateral panels show different time delays.

**DEPENDENCE OF THE PHASE DIFFERENCE AND FREQUENCY VARIANCE ON THE STRENGTH OF THE INTRA-NETWORK COUPLING.**

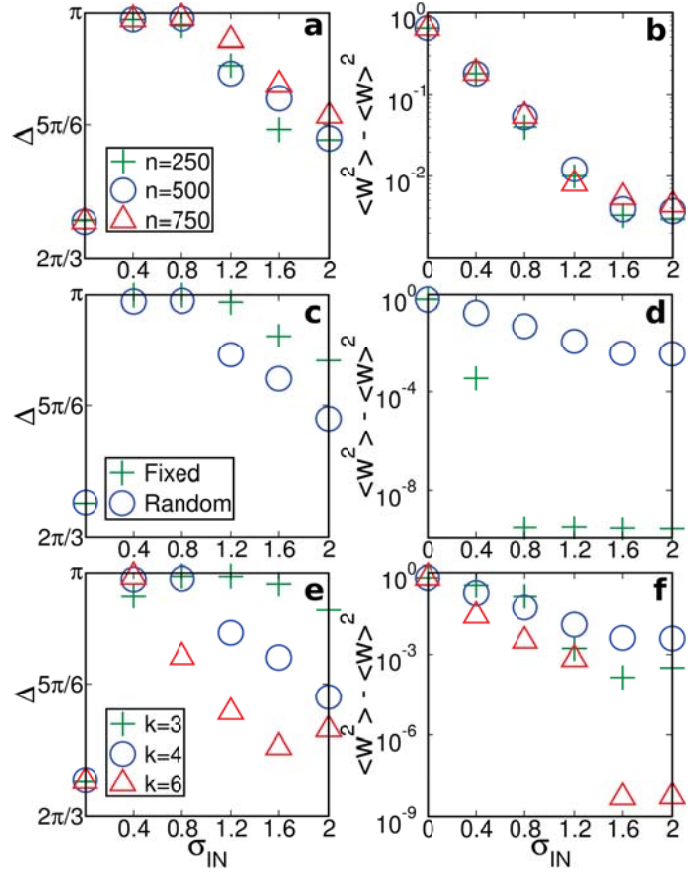

FIG. S2: Panels on the left column represent the phase difference $\Delta$ between pairs of inter-network neighboring oscillators. Panels on the right column represent the average variance of the frequency of oscillators in the same network. Panels a) and b) are results for networks with different sizes and average degree of four. Panels c) and d) are results for networks with node degree fixed at four in comparison with networks with randomly distributed degree of average four. Panels e) and f) are results for networks with diferent average degrees. Each point in all plots represents an average over 500 samples with $\omega_0 = 2.75$, $\tau = 1.53$, and $\sigma_{\text{EX}} = 1.5$. Panels c)-f) represent results for networks with 500 nodes. The standard deviations in all cases are smaller than the symbols.

# FREQUENCY COMMUNITIES IN TRANSITION REGIONS

We analyze the transitions between synchronization regimes shown in the diagram of Fig. 3 in the main text. The transitions between regions of one and two stable frequencies are not smooth (See Fig. S3). We investigate three regions of the parameter space where transitions between states are expected. All are consistently abrupt, i.e., a small difference in the coupling strength triggers a bifurcation in the possible stable frequencies. Differences in the transition point can be explained in the context of finite size effects. A certain combination of coupling strengths and time delays leads to synchronization at a single stable frequency, but a small difference in any of the parameters can give rise to two stable frequencies, with the possibility of breathing synchronization, and a strong dependence of the final state on the initial conditions.

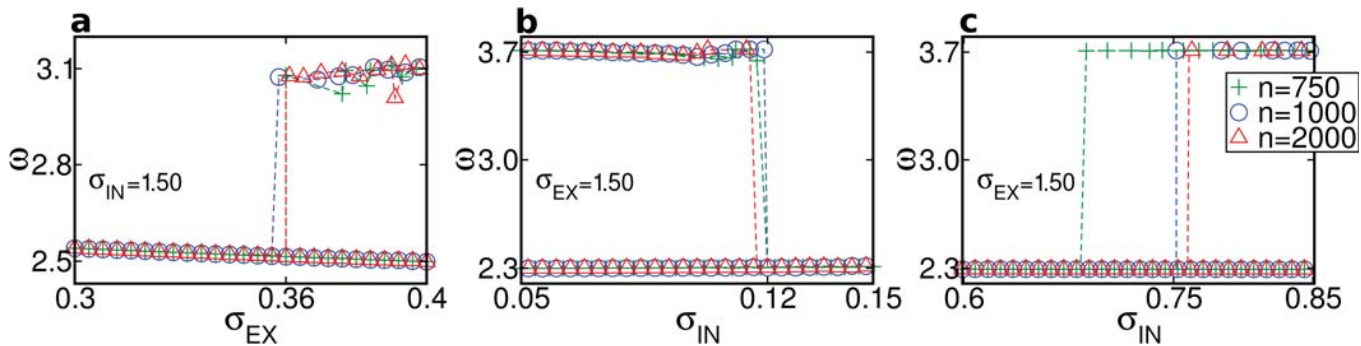

FIG. S3: For certain combinations of coupling strengths, one can observe abrupt transitions in which global synchronization is lost (Panels **a** and **c**) or recovered (Panel **b**). Points are averages over oscillators in each frequency community. Simulations were performed with 300 samples with $\tau = 1.53$ and $\omega_0 = 2.75$. In Panel **a**, we show results for $\sigma_{IN} = 1.5$, while Panels **b** and **c** are for $\sigma_{EX} = 1.5$. Panels a)-c) are averages over 300 pairs of random networks of 1000 nodes, 300 pairs of 1500 nodes, and 300 pairs of 1950 nodes. The standard deviation in all cases are smaller than the symbols. To reduce noise, we consider only frequencies with a relative occurrence of more than 10% to calculate averages.


\end{document}